# An overview of metrics-based approaches to support software components reusability assessment

Miguel Goulão, Fernando Brito e Abreu

Informatics Department

Faculdade de Ciências e Tecnologia / Universidade Nova de Lisboa

Quinta da Torre, 2829-516 Caparica, Portugal

{miguel.goulao|fba}@di.fct.unl.pt

http://ctp.di.fct.unl.pt/QUASAR/

## Abstract

**Objective**: To present an overview on the current state of the art concerning metrics-based quality evaluation of software components and component assemblies.

**Method**: Comparison of several approaches available in the literature, using a framework comprising several aspects, such as scope, intent, definition technique, and maturity.

**Results**: The identification of common shortcomings of current approaches, such as ambiguity in definition, lack of adequacy of the specifying formalisms and insufficient validation of current quality models and metrics for software components.

**Conclusions**: Quality evaluation of components and component-based infrastructures presents new challenges to the Experimental Software Engineering community.

**Keywords:** Component-Based Software Engineering; Component Evaluation; Software Metrics; Software Quality.

## 1 INTRODUCTION

### 1.1 Motivation

Component-based development (CBD) is playing an increasing role in the software industry [1, 2]. There is an economic push to such growth: the claim is that CBD allows the reduction of cost and time to market, while increasing software quality, through reuse [3]. The rationale is that cost savings can be obtained through economy of scale, while improved quality results from the reuse of such components in different environments and applications. Recently, a component broker conducted a case study with the cooperation of component producers [4]. Its goal was to estimate the return on investment of commercial-of-the-shelf components (COTS). The referred case study reports that the costs of acquiring such components are about 1/50 of the ones for developing their required functionalities from scratch.

The Software Engineering Institute (SEI) defines a component as *"an opaque implementation of functionality subject to third party composition and conformant to a component model"* [5]. With an increasing percentage of component-based architectures relying on black-box software components, the quality of such architectures depends, to a large extent, on the quality of those components and on the interactions among them [6]. Therefore, components evaluation should be integrated in CBD [7].

One of the key roles in CBD is that of the component assembler. A component assembler starts with application requirements, searches component repositories for selecting appropriate components, and assembles them by providing the required glue [3]. His focus of attention is on component composition rather than on component construction. From a component assembler perspective, being able to assess the complexity of candidate alternative component assemblies is crucial. This task is difficult, as he has to consider the integration of components that may be acquired from different providers, each offering a different documentation detail level for every component. Deciding whether to reuse components or to develop the corresponding functionality from scratch is also part of the tasks performed by the component assembler.

In this context, it would be helpful for a component assembler to have an objective, integrated, independent view of existing techniques that may assist him in this task. Objectivity can be obtained by performing a quantitative comparison among alternatives, rather than just a qualitative one. The integration of such comparative studies can be facilitated by using a common evaluation framework. Independency can only be achieved by integrating assessments from different independent sources. In contrast, it is common for component assemblers to be forced to base decisions on their personal experience and the qualitative judgement of "experts". Component assemblers are therefore potentially vulnerable to biased information sources and hype.

## 1.2 The need for CBD-specific evaluation techniques

Building upon SEI's definition of a component, we can contrast evaluation of CBD with that of object-oriented or structured development. The first major difference relates to the opaqueness of components. While several metrics-based approaches for evaluation of software complexity (e.g: McCabe metrics) rely on access to the source code, similar approaches for CBD should depend only on the information publicly available on black-box components. Indeed, the component's source code is often not available to component assemblers. Moreover, there is a problem of scope. The component assembler is not concerned with the internal complexity of a component, but rather with the complexity involved in reusing it. Internal code metrics for analysing the components are not useful, from his point of view. Instead, complexity analysis on the interface of a component, the contracts associated with it and the adaptability of the component to different contexts should be assessed.

There is no widely accepted quality model for CBD, although the community has proposed extensions of the ISO9126 standard [8] to fit the needs of CBD [6, 9]. Such a model is required for quality evaluation, whether this evaluation is of a qualitative or quantitative nature. A typical example of qualitative evaluation is an expert's opinion on the component artefact. Qualitative evaluation is subjective, posing problems in results comparison and generalization. Besides, experts may not be available at all. The quantitative approach to evaluation provides a more pragmatic way of dealing with this problem. It consists on defining, collecting and analyzing objective quantitative metrics that can be used, when framed by a quality model, to replace (or complement) the expert's opinion in an automated fashion. The goal is to provide heuristics-based help as guidance to practitioners in the component selection process.

## 1.3 Overview outline

Our overview is focused on proposals for metrics-based approaches to component reusability assessment. To the best of our knowledge, there is a lack of comparative reviews of such proposals. Our contribution in this paper is threefold: we define an evaluation framework for metrics proposals; we use that framework to provide a comparative study on component reusability evaluation proposals; and we outline an approach to support the replication of experiments to assess such proposals.

This overview is organized as follows: in section 2 we discuss typical problems in metrics-based approaches to software products assessment; in section 3, we present a framework for our review on the state of the art of metrics-based assessment for reuse in CBD; in section 4 we discuss different metrics-based approaches to component evaluation, both in isolation and in assemblies; our view on how the state of the art may be improved is presented on section 5; finally, conclusions are presented in section 6.

# 2 COMMON SHORTCOMINGS OF METRIC-BASED APPROACHES

## 2.1 Metrics and their underlying quality model

The lack of a widely accepted quality model for CBD is the first challenge for a component assembler in his component selection process. There are some proposals of quality models for CBD, such as [9], where an adaptation of the ISO9126 [10] for component software is proposed, but none of these proposals have achieved an industry-wide acceptance, yet.

Often, metrics definition is not performed to meet the information requirements of a particular quality model, but rather in an ad-hoc fashion. In the absence of such a reference model, interpreting measurements is troublesome. Consider the example of Lines of Code (LOC) measurement, which could be used in the assessment of white box components. If we simply define how to count the LOC with no reference to how we plan to use them, we are in fact only defining a measurement, but not a metric. Defining the latter implies referring to a framework (the quality model) upon which we plan to interpret the measurements. The LOC measurement has been used in several contexts. As a size (or complexity measure), LOC has been used, among other things, to assess the maintainability of software and the productivity in developing code. Each of these usages requires a different validation process that should provide an answer to the following questions:

- How does size influence maintainability?
- How does size influence developer productivity?

There is no lack of controversy regarding each of these questions. Factors such as the code reuse level, the particular kind of reuse, or the coding style, may have a significant impact on the value of LOC. Our point is that it is not possible to define and validate a metric, without clearly stating what is its intended usage, and that implies specifying the underlying quality model.

## 2.2 Metrics ill-definition

The metrics ill-definition problem occurs due to several reasons. Metrics definitions are often presented without the corresponding context. Without expressing which is the underlying metamodel upon which the concepts and their interrelationships are defined, metrics definitions become subjective, as different interpretations on which those concepts are and on how to perform the measurements are possible. Finally, metrics definitions are performed without an underlying formal approach that uses the previously mentioned metamodel as contextual input. The formal specification of metrics should address not only how the underlying concepts are accounted for and their interrelationships are traversed to collect the required metrics, but also the pre-conditions that must be met to allow the collection of such metrics.

Without clear and precise definitions of metrics, it may be impossible to consistently develop tools to collect those metrics, or to discuss their properties in a mathematically sound way. The usage of natural language is a typical metrics definition problem. One of the first books on metrics for object-oriented design contained natural language definitions for all its metrics [11]. While this may be considered helpful as a first glimpse on the metric's objective, the absence of a formal definition may hamper its systematic and repetitive collection and validation by different researchers or practitioners. Consider the following natural language definitions, borrowed from [12]:

- *"Component Interface Complexity Metric (CICM): Component interface complexity metric should provide an estimate of the complexity of interfaces. Such a metric could be helpful in improving the systems quality because complex interfaces complicate the testing, debugging and maintenance."*

- *"Component Resource Utilization Metric (CRUM): Resource utilization metric should measure the utilization of target computer resources as a percentage of total capacity."*

The first definition is a typical example of a "wish list" metric proposal. Although it contains an intuition to the authors' intentions when defining it, the description is too vague with respect to how the interfaces complexity should be measured. The second definition is more objective, in the sense that it implies that the metric is defined as a ratio between used and available resources. It completely omits which resources should be measured and how they could be measured. For the sake of argument, assume that we wish to instantiate the second metric by computing CRUM considering memory as the resource under scrutiny. Which would be the conditions for performing the measurement? Should we consider the average memory used by the component during its lifetime, its highest value during a particular period of usage, or any other option? Should we consider the total physical memory of the target computer as a baseline, or discount the memory used by other applications, namely the operating system being run by that computer? There are too many points of uncertainty in this kind of definition, leading to points of variation in the implementation of tools for collecting them.

Note that even the apparently trivial LOC definition as a count of lines of code is susceptible to different interpretations, in part due to its vulnerability to coding style options. When analyzing a source code file, should we make a simple count of lines, or should we omit, for instance, blank lines? Should comment lines be counted as well, or omitted? How do we deal with text wrapping? Should we pre-process the source code to ensure a uniform formatting style?

Although in principle one can always detail all the counting rules down to their most intricate details, natural language definitions of metrics are often incomplete and ambiguous. A consequence is that different tools collecting allegedly the same metric may provide different values for that metric, while analyzing the same artefact. This hampers the comparability of metrics collected by independent teams using different tools. Results interpretation may also be flawed, due to these potentially different interpretations of the natural language definitions.

A common approach to increase the quality of metrics specifications is to use a combination of set theory and simple algebra to define metrics. Consider the following example, borrowed from Hoek *et al.* [13], for the provided (*PSU$_x$*) service utilization metric.

$$PSU_x = \frac{P_{Actual}}{P_{Total}}$$

Hoek *et al.* define a service as follows: *"Under the term service, we include such things as public methods or functions, directly accessible data structures, and any other kind of publicly accessible resource one may be able to express in an ADL."* Their intention is to define these metrics in a generic way, so that they are not tied to any particular ADL (Architecture Description Language), or service. The price to pay for this option is that different implementations will consider different kinds of services as relevant. Although the metrics formulas are objective, the selection of the elements to be included in such formulas is ambiguous, making these metrics ill-defined.

The alternative is to use a formal approach to define metrics. Dumke *et al.* proposed a taxonomy for formal approaches to software measurement [14], as well as a discussion on their strengths and weaknesses. The categories include algebraic, axiomatic, functional, rule-based, structure-based, information-theoretical, and statistical approaches. In our opinion, their taxonomy lacks cohesion, in the sense that some of the former 6 groups of approaches concern metrics definition, while the latter (statistical) is mostly concerned with how to treat data obtained from those metrics. The main shortcoming of formal approaches to metrics definition is that understanding them requires mathematical skills that are often not held by common practitioners.

## 2.3 Insufficient validation of metrics-based approaches

This problem is not specific to component-based development. As pointed out in [15], Experimental Software Engineering research, in general, tends to be fragmented and not properly integrated. This leads to the absence of a culture of replication of experiments and of systematic reviews of the existing approaches, like, for instance, is common practice for medical researchers [16]. This shortcoming of current Experimental Software Engineering practices has been identified in several surveys. A systematic review on controlled experiments conducted in software engineering [17] has reported that, out of 5453 scientific papers published in 12 leading software engineering journals and conferences from 1993 to 2002, only 103 (1.9%) of them reported controlled experiments in the realm of software engineering tasks. These included only 14 series of replications, where controlled experiment replications were performed. Only 6 of these series replications included replications performed independently (not by the original authors). While the above mentioned review focused on controlled experiments, its observations are consistent with those of other surveys concerning alternative forms of evidence-based validation of software engineering claims, ranging from controlled experiments to observational studies (e.g. [18, 19]). The lack of standard protocols[1] to conduct experimental work in software engineering is one of the problems undermining the availability of evidence to support decisions such as those that have to be made by component assemblers, although there are recent guidelines proposals aimed at mitigating this problem [20].

## 3 A FRAMEWORK FOR CHARACTERIZATION OF PROPOSALS

It is useful to have a common framework, upon which we can characterize the reviewed work. Framework-based analysis fosters a more systematic approach to proposals assessment than the one usually achieved through a more traditional, non-structured, narrative review. Evidence collected in the realm of medical sciences show that narrative reviews tend to lead to more informal and subjective methods to collect and interpret the studies and even to selective citation of literature to reinforce preconceived notions [16]. In contrast, having a framework for characterizing proposals fosters a more objective analysis, partially mitigating the shortcomings of narrative reviews. The framework also helps readers identifying which proposals are likely to be applicable to their own context, and which are not. Therefore, we propose here a framework upon which we will base our review.

This framework includes a set of qualitative characteristics plus a quantitative assessment scheme, based on ordinal scales. The quantitative assessment enforces the required comparability of proposals. Together, the qualitative and quantitative parts provide a basis for identifying the strengths and shortcomings of each proposal, as discussed in the previous section. The first four items of this structure aim to provide a very brief overview of the proposals, while the last aims to characterize each proposal according to its maturity level.

- **Scope** – This refers to the granularity level and type of artifacts that are the target of the metrics-based assessment proposal. A typical contrast is between coarse and fine-grained components. Another one is that while some components are white-box, others are black-box. The scope definition constrains the assessments that can be performed on components.

- **Intent** – A description of the main objectives of the proposal, to help the reader assessing the extent to which each approach may help achieving those objectives.

---

[1] In the context of experimentation, a protocol is a specification of the steps to be followed while conducting an experiment, from the experiment setup to data analysis. Following standard protocols increases the comparability of individual studies, as it fosters homogeneity in the data collection process.

- **Technique** – This refers to how the metrics were defined and validated. The metrics definition technique may range from a purely informal description to a formal definition. Several forms of validation of the proposals may have been attempted, both by the metrics proponents and other researchers and practitioners. In metrics proposals, validation efforts range from case-studies that use toy examples and aim at illustrating the metrics definition and collection, to series of controlled experiments performed with real-world examples.
- **Critique** – Here, we provide a qualitative assessment of the most noticeable features of the proposal, including its most interesting aspects, as well as its main shortcomings.
- **Maturity** – The maturity level of the proposal provides a comparison framework based on the usage of ordinal scales to characterize the metrics proposals according to four different dimensions: the underlying quality model, the mapping quality between metrics and the quality model, the formality of the metrics definition, and the extent to which the proposal was validated.

To assess the maturity of the proposals, we start by identifying a set of rating scales concerning different aspects of metrics-based quality evaluation. For each of those rating scales, we then identify several levels of maturity that will aid us in the graphical depiction of proposals maturity. Table 1 presents a condensed view of our maturity comparison framework.

| Maturity level | Quality Model (**QM**) | Mapping Quality (**MQ**) | Metrics definition (**MD**) | Level of Validation (**LV**) |
|---|---|---|---|---|
| 0 | N/A | N/A | N/A | N/A |
| 1 | Ad-hoc | Ad-hoc | Wish list | Anecdotal |
| 2 | Structured | Rationale | Informal | Small experiment |
| 3 | Uncorrelated | Goal-driven | Semi-formal | Industrial experiment |
| 4 | Validated | Validated | Formal | Independent |

Table 1 - A metrics proposal maturity comparison framework

The maturity level is of an ordinal nature, ranging from 0, where the dimension is not available in the proposal (N/A in all rating scales), to 4, where the proposal has reached a high maturity level. It should be noted that a proposal's maturity does not necessarily reflect its potential interest. For instance, a radical proposal in an emerging field may be promising, while not yet evidencing high values across all the aspects of our comparison framework. On the other hand, we will expect that within a reasonable period of time, the same proposal will mature.

In the next section, we will present several proposals for metrics-based assessment of reusability in CBD. For presentation purposes, we will use the following maturity mask, where **level** is replaced by the appropriate value for each proposal:

> QM[level]; MQ[level]; MD[level]; LV[level]

The **Quality Model (QM)** represents the extent to which the metrics proposals fit into a quality model. For the Quality Model, the identified categories, by increasing level of maturity, represent:

0. **N/A** – The proposal is not explicitly related to a quality model.
1. **Ad-hoc** – A set of quality characteristics are identified.
2. **Structured** – Quality characteristics are organized, typically in a hierarchy.

3. **Unassociated** – Quality characteristics are shown to be independent, to avoid assessing the same quality aspect repeatedly.
4. **Validated** – The quality model is conveniently validated through experiments.

The **Mapping Quality (MQ)** represents the level of integration between the model and the metrics which are chosen to assess quality based on that model. The represented categories are:

0. **N/A** – Metrics are not related to a quality model.
1. **Ad-hoc** – Metrics are mapped to quality attributes in an ad-hoc fashion.
2. **Rationale** – A discussion on the rationale of the mapping is provided.
3. **Goal-driven** – Metrics are defined to answer specific evaluation needs, following an approach such as the Goal Question Metric [21].
4. **Validated** – Building on the previous level, metrics are shown to effectively fulfill the specific evaluation needs raised by the quality model.

Concerning **Metrics Definition (MD)**, we use the following categories:

0. **N/A** – The proposal is only qualitative.
1. **Wish list** – The authors informally identify the need for a certain kind of metrics, without actually proposing any.
2. **Informal** – A natural language description of the metrics is provided by the authors.
3. **Semi-formal** – Some degree of formalism is used in the metrics definitions. Typically, the metrics themselves are defined through mathematical expressions, but the underlying concepts being measured are only informally specified.
4. **Formal** – A formal definition of the metrics based upon the underlying concepts is provided.

Finally, the **Level of Validation (LV)** is classified according to the following categories:

0. **N/A** – The proposal does not include any example of metrics collection.
1. **Anecdotal** – Anecdotal examples are provided to motivate the usefulness of the proposed metric. Sometimes, they are complemented with some descriptive statistics.
2. **Small experiment** – An experiment is carried out to assess the metrics, with some statistical approach to analyze the collected data, but the sample of analyzed artifacts does not allow inference (conclusions generalization beyond the sample used in the experiment).
3. **Industrial experiment** – An experiment with a significant sample of artifacts is carried out, with real-world artifacts and adequate statistical analysis.
4. **Independently validated** – Experiments conducted by independent research teams confirm the original proponent's claims.

## 4 METRICS FOR REUSE IN CBD

Our overview focuses on metrics-based approaches that aim at helping component assemblers to choose adequate components. The selected proposals share a concern for assessing, somehow, the reusability of components. For easier reference, the proposals will be identified by the name of their first author, both in their textual description and in the chart with the overall comparison, presented in Figure 1. A reference to the corresponding papers is provided on the top of each of the proposals review.

We have divided these metrics into two groups. The first one contains proposals that consider the components in isolation. The second relates to proposals that attempt to help assessing components in a given context, which is typically either a component assembly or a component library.

### 4.1 Approaches to the evaluation of individual components

**Bertoa's quality model and metrics** [9, 22, 23]

| | |
|---|---|
| **Scope** | COTS software |
| **Intent** | To introduce a quality model as an adaptation of the ISO9126 for component-based development [9]. The adaptation of the ISO quality model consists on assuming that the software will include black-box components and change the quality model accordingly, so that any assessment of reused software takes into account this restriction. A set of metrics to assess the attributes of that quality model is also proposed. Its rationale is that the metrics collection has to be defined considering the information made available by component brokers. While the first attempt at metrics definition covers transversally the quality model, more recent work by the same authors focuses on the usability of components, as perceived by component assemblers [22]. |
| **Technique** | Although some of the metrics definitions included mathematical formulae, most definitions were informal [9, 22]. In [23], where a validation effort for metrics concerning the usability of components is presented, all metrics definitions are presented in natural language. This presentation is complemented by a metamodel describing the information available from COTS vendors that concerns usability. The metrics set includes metrics on not only the COTS components, but also on their documentation. The metrics collection requires a strong manual intervention, as several of the metrics are collected from the analysis of the available documentation of COTS components. The validation was conducted in a series of 5 experiments (one of them was a replica conducted by peers) with a total of 68 subjects that were asked to evaluate a sample of 12 COTS components. The first three experiments concerned a subjective analysis performed by the participants on each of the components in the sample. The remaining two experiments consisted on an assessment of component reusability through the analysis of the performance of users while answering objective questions concerning the availability of specific tasks and services in the components that made up the sample. Subject's performance was measured as a combination of correctness of responses and time required for providing such responses, and was used as an indirect measure of component reusability. |
| **Critique** | By using the information made available by vendors, there are limitations concerning the ability to automate metrics collection, due to the noticeable lack of standards in data presentation by COTS producers and brokers. To overcome this problem, a UML model for the classification of COTS usability is proposed, but populating that model in an automated fashion remains an open challenge. From the original set of metrics [9], some were dropped out due to difficulties in their collection. With respect to the validation efforts, the proponents' attempt to build up a set of experiments was successful in what concerns the replication of the experiment by an independent team, but the small component sample is probably the most noticeable threat to validity of the experiment series. |
| **Maturity** | QM [Structured]; MQ[Rationale]; MD[Informal]; LV[Small experiment]. |

**Gill's quality attributes** [12]

| | |
|---|---|
| **Scope** | Black-box components |
| **Intent** | To propose a set of guidelines on how to select metrics for black-box components. |
| **Technique** | No actual metrics are defined. Instead, the authors informally present a set of quality attributes that should be evaluated through metrics. |
| **Critique** | The proposal includes an interesting discussion on the focus shift caused by the specificity of black-box components evaluation, as opposed to the evaluation of OO design, or structured software and provides an interesting roadmap for research in metrics-based component evaluation. |
| **Maturity** | QM[Ad-hoc]; MQ[Rationale]; MD[Wish list]; LV[N/A]. |

**Dumke's metrics for reusability of JavaBeans** [24]

| | |
|---|---|
| **Scope** | White-box Java Beans |
| **Intent** | To present a metrics set for reusability of JavaBeans. |
| **Technique** | Informal definition of metrics, relying on access to the source code. The metrics in this set are adapted from other contexts, such as OO design (e.g. percentage of public methods) and structured programming (e.g. maximal McCabe complexity number, for a method in the JavaBean class). |
| **Critique** | The white-box view of components renders this approach inadequate for evaluation by independent component assemblers. The internal complexity of a component method should not be relevant for the understandability of its interface and the component's reusability. |
| **Maturity** | QM[N/A]; MQ[Ad-hoc]; MD[Informal]; LV[Anecdotal]. |

**Boxall's interface textual complexity metrics** [25]

| | |
|---|---|
| **Scope** | Interfaces of components developed with C, C++, Java or Eiffel. |
| **Intent** | To define a set of metrics to assess interface complexity, measuring aspects of components' interfaces, such as the interface size, number of distinct arguments in operations, level of repetition of such arguments, the commonality in identifiers, identifier's length and the density of reference arguments. |
| **Technique** | Metrics are defined through a set of mathematical expressions, but the elements of such expressions are informally described. |
| **Critique** | The level of detail in the analysis of arguments in the interface is richer than in other approaches, in what concerns the relevance of naming conventions for component interfaces' understandability. However, this approach does not address other potentially interesting aspects in the interface, such as arguments' complexity. |
| **Maturity** | QM[Informal]; MQ[Rationale]; MD[Semi-formal]; LV[Small experiment]. |

**Washizaki's reusability metrics for black-box components** [26]

| | |
|---|---|
| **Scope** | JavaBeans Interfaces. |
| **Intent** | To propose a metrics set for assessing the reusability of JavaBeans. The metrics set is defined in the scope of a quality model for black-box component reusability, considering understandability, adaptability and portability as relevant sub-characteristics. More refined criteria are then defined for each of these sub-characteristics, as well as metrics to assess JavaBeans in light of such criteria. |
| **Technique** | Metrics are defined as ratios of the effective use of a given sort of interface feature (e.g. BeanInfo class, readable properties, writable properties, methods with parameters and methods with no return value) when compared to its potential use. |
| **Critique** | It can be argued that the analysis of the interface complexity is over-simplistic since at least two aspects are not considered: (i) the complexity of arguments, and (ii) the repetition of argument types. In both cases no distinction is made. Intuitively, increased complexity and variety of argument types would decrease the understandability of the component's interface. |
| | Washizaki's metrics set was validated with a case study where the reusability of over 120 components was assessed, both with this metrics set and by a panel of experts. Results show a high correlation between both assessments, indicating that the metrics defined in this set can indeed be used to assess component's reusability. However, an independent case study showed the metrics to be unreliable for components with a small number of features on their interface [27]. Further independent analysis is still required. |
| **Maturity** | QM[Structured]; MQ[Rationale]; MD[Semi-formal]; LV[Industrial experiment]. |

**Gill's interface complexity metrics** [28]

| | |
|---|---|
| **Scope** | Black-box components' interface |
| **Intent** | Besides the complexity aspects of interfaces' signature, this proposal also considers constraints upon those interfaces, as well as their packaging, to account for different configurations that the interface may present, depending on the context of use. |
| **Technique** | The overall complexity is defined as the weighted sum of the complexities related to signature, constraints and packaging of the interfaces. For each of these aspects of interface complexity, a definition is also proposed, again using weighted sums of features (e.g. events and operations count, for signature's complexity). |
| **Critique** | Although Gill's proposal has the merit of including constraints and packaging complexities on the assessment, it still lacks any sort of empirical assessment. This hampers the ability of the authors to assign values to the coefficients on their definitions, and, more significantly, our ability to assess the extent to which this approach helps common practitioners to choose among alternative components. |
| **Maturity** | QM[N/A]; MQ[N/A]; MD[Informal]; LV[N/A]. |

## 4.2 Approaches to the evaluation of component assemblies

The approaches described in the previous section are mostly targeted at the assessment of components in isolation. They rely on the assumption that the quality of software components influences in some way the quality of the assembled system. The apparent conclusion of this would be that a component assembler should always try to choose the best components in order to optimize the quality of the assembled system. This may reveal to be naïf, since we should also consider the context in which the component will operate. Determining how well a component integrates with other components in an assembly may lead to an evaluation that is more worthy to the component assembler than the one made in isolation [29]. This change of scope allows the component assembler to focus on the quality for his target product: the component assembly.

**Sedigh-Ali's quality characteristics** [30]

| | |
|---|---|
| **Scope** | COTS |
| **Intent** | To discuss the requirements for metrics for CB-architectures based on relevant quality aspects. The authors also present a taxonomy on the categories of costs related to software quality, with cost drivers such as quality improvement, low quality prevention, software failures and external costs related to those failures. |
| **Technique** | High level discussion, rather than a concrete proposal. |
| **Critique** | The main contribution of this paper is an interesting discussion on requirements for metrics for CB architectures, measured at a system level, including insights on how to choose relevant metrics. However, this is an exploratory work based on expert opinions alone, rather than on some sort of quantitative evidence to back up the presented arguments. |
| **Maturity** | QM[Ad-hoc]; MQ[Ad-hoc]; MD[Wish list]; LV[N/A]. |

**Seker's coupling and cohesion for CBD** [31]

| | |
|---|---|
| **Scope** | Black-box components and component assemblies |
| **Intent** | To define coupling and cohesion metrics for CBD. |
| **Technique** | The metrics are defined using an information theory based approach where components and component infrastructures are represented as graphs. |
| **Critique** | This approach adapts the well-known concepts of coupling and cohesion to the scope of CBD. Except for the nodes in the graph being black-box components rather than classes, the proposal is similar to coupling and cohesion for OO design. |
| **Maturity** | QM[N/A]; MQ[N/A]; MD[Semi-formal]; LV[N/A]. |

**Hoek's service utilization metrics** [13]

| | |
|---|---|
| **Scope** | Software product lines |
| **Intent** | To propose a metrics set that allows assessing software product lines based on service utilization. The rationale for their need is that service utilization in product lines implies a degree of optionality among the components that get used in a given configuration. While some services and components will be part of all configurations of that product line, others are optional. Structural variability is also an issue, as the component assembler has to choose among a range of alternative configurations. Product lines are also typically hierarchical, composed of a set of components, each of which with its own internal structure. The combination of the above mentioned constraints violates the assumptions of most structural metrics that the system structure under evaluation is: (i) single - no optionality considered, snapshots of the system are usually evaluated); (ii) fixed - no structural variability, the system structure is assumed to be kept constant throughout the evaluation; and (iii) flat - the implications of the hierarchical decomposition of the system are not considered in the metrics definition. |
| **Technique** | The metrics are defined around the concept of service utilization (the rate of usage of provided and required services of a component). For individual components, metrics are simply ratios of used services (both for required and provided ones), whereas for component architectures which are fixed and flat (assemblies) these ratios are obtained using the sum of used services against the total of available services. |
| **Critique** | Of all the proposals presented in this overview, Hoek's is the one that best fits the notions of architectural components and assemblies' evaluation rather than individual components' evaluation. |
| **Maturity** | QM[N/A]; MQ[N/A]; MD[Semi-formal]; LV[Anecdotal]. |

**Inoue's ranking significance [32]**

| | |
|---|---|
| **Scope** | Software component libraries. Although the proposal is instantiated to Java class libraries, it is generic and could be used with other sorts of components, from fine to coarse-grained, both white and black-box. |
| **Intent** | To enable the implementation of a Java class retrieval system (SPARS-J) that aids developers finding out relevant classes for reuse through natural language queries. As the results of those queries tend to be too broad, a ranking system is required to sort the search results in a convenient fashion. The approach is inspired by the computation of impact factors of scientific publications (research papers, books, etc.) and the ranking mechanisms used by modern web search engines. Components are ranked with respect to their reuse in an existing software baseline. The most reused components have a higher rank and are thus presented at the top of query results, as they are more likely to be of interest for the practitioner performing the query. |
| **Technique** | The component rank model uses a weighted directed graph representation for software components, where nodes represent the components and edges represent the use relationships among those components. The weight of each node is computed as a function of the weight of its incoming edges. In turn, the weight of each edge is computed as a function of the weight of its origin node and the number of outgoing edges that node contains. The computation of all these weights corresponds to obtaining a stationary distribution of the Markov chain [33] that the underlying graph models. |
| **Critique** | One of the most noticeable features of this approach is that reuse is assessed in terms of the effective reuse of software components, rather than in terms of expected reuse (e.g. predicted from the component interface's characteristics). This means that the metrics are useless from the point of view of a component developer. In turn, they may be very useful for component assemblers, as they help locating the most frequently reused components. From all the presented proposals, this was clearly the most thoroughly validated one. The ranking system is in use in two different companies, where a small case study concerning user satisfaction with the ranking system was conducted. The results were very encouraging, although a larger sample of users would be required to confirm them. More important, the ranking system was tested with a set of about 6100 components, from the JDK 1.4.2, on a first observational study, and 180000 components from publicly available Java component libraries, collected from SourceForge.net, on a second one. In both cases, the ranking system obtained significantly better results than those of non-specialized search engines. The authors do not specifically |

present the underlying quality model, although the proposal assumes that leveraging software component reusability is a promising approach to the development of high-quality software.

**Maturity**     QM[N/A]; MQ[N/A]; MD[Semi-formal]; LV[Industrial experiment].

## 4.3 Lessons learned

Figure 1 represents an overview of the maturity levels of each of the proposals described on previous sections. In this chart, from left to right, we present each proposal; from front to back we present each of the analysed rating scales. On the vertical axis we have the maturity level, as defined in Table 1. The overall low level of maturity throughout the several rating scales supports the claim that research in the area of software components quantitative evaluation is still on a very early stage. We can revisit now the three aspects highlighted in section 2.

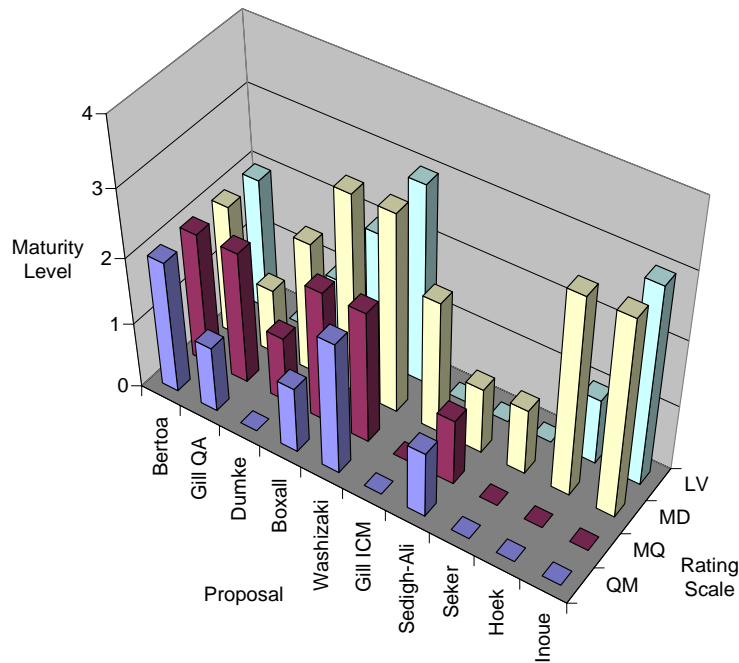

Figure 1 – *Overall proposal maturity assessment*

### 4.3.1 Lack of an underlying quality model

This shortcoming is related to the generally weak relationship among metrics proposals and quality attributes. In the best-case scenario we found proposals where a structured quality model was included, along with a discussion associating the metrics with the quality attributes defined in the model, including the expected effect that variations in those attributes may have on metrics. Washizaki and Bertoa's works were the ones dedicating more attention to this problem, while several other proposals do not explicitly address it, in the reviewed publications. This shortcoming of metrics proposals follows a more general tendency observed in other contexts, such as that of OO development, where metrics proposals often lack an adequate quality model context [34].

### 4.3.2 Metrics ill-definition

None of the reviewed proposals includes a formal definition of metrics. In some cases, the author's intentions were clearly to leave the metrics definitions abstract enough for readers to adapt those definitions to their own context (e.g. Hoek's metrics). There is a fairly balanced distribution between whish-lists (3), informal definitions (3) and semi-formal definitions (4) of metrics. Since the majority of definitions are too informal, replicated experiments aimed at validating these proposals are bound to have difficulties related to the tacit knowledge problem: insufficient information provided by the original authors of an experiment causes difficulties in its replication. In this case, the tacit knowledge concerns the definition of the metrics, where non-stated assumptions may lead to different interpretations of the original metrics definitions. While the tacit knowledge problem, as describer by Shull *et al.* [35] is wider (it refers to all relevant information for replicating an experiment, from its requirements statement to the results packaging, which is not clearly specified in the experiment reporting, leading to possibly wrong assumptions by those who attempt to replicate the experiment, with respect to what was really done in the original experiment), it could be mitigated, in what concerns metrics definition, by providing a formal definition of all the defined metrics.

### 4.3.3 Insufficient validation

Insufficient validation occurs when independent cross validation is not performed, mainly due to difficulties in experiment replication. Independent metrics validation (not performed by their authors) is fundamental for their proof of usefulness before widespread acceptance is sought.

It is worth noticing that only Washisaki's and Inoue's proposals were validated with industry-level observational studies. Inoue's validation efforts included two case studies carried out in different companies and used significantly larger samples than any other proposal. It is fair to recognize their validation efforts level as well above the usual state of practice with software metrics, both in the context of CBD metrics and otherwise. The validation efforts on Bertoa's proposals were also noteworthy for their emphasis on replication, but their main shortcoming seems to be that their metrics collection is partly manual. The majority of the proposals discussed here were not validated at all.

## 5 MITIGATING SOME OF THE IDENTIFIED PROBLEMS

### 5.1 Providing adequate context for metrics proposals

Metrics proposals should be framed in the context of a quality model, to prevent the collection of data for which there is no expected usage, with the corresponding waste of valuable resources. The quality model should guide the establishment of goals, for which research questions would be made, leading to the definition of objective metrics to answer those questions. Although there is a well-known and widely accepted approach named Goal-Question-Metric[21] that aims at guiding the definition of software metrics, the results of our survey showed that the community is still not using this approach as much as would be desirable.

### 5.2 Facilitating the replication of validation efforts

Automated metrics extraction is fundamental to foster independent validation efforts. Manual collection of metrics has been shown to be error prone and vulnerable, for instance, to the lack of adherence to sound, widely accepted, coding principles [36]. Furthermore, the effort required for large scale manual metrics extraction is prohibitive. Unfortunately, to the best of our knowledge, most of the proposed metrics were only tested by their authors, using proprietary or experimental, non-publicly available, tool support, therefore limiting Experiments replication. This limits knowledge sharing, both in the research and practitioner's communities, hampering results comparison. We have proposed elsewhere an approach to mitigate this problem [27, 34, 37, 38], which relies on the usage of a metamodel to formally define the concepts we aim to measure, and Object Constraint Language (OCL) expressions to define metrics over that metamodel. It can be summarized as follows:

- Selecting or producing an adequate metamodel describing the domain concepts (for instance, for CB systems, we could use the CORBA Components Metamodel[39], or an extract of the UML 2 metamodel corresponding to component diagrams[40, 41].

- Specifying the metrics using OCL [42] upon the previous metamodel. Notice that the latter is specified as a UML class diagram that can be traversed using OCL expressions.

- Instantiating the metamodel, with meta-objects and meta-links corresponding to the target software piece (e.g. code, or model element) that we want to measure.

- Collecting the metrics using an OCL-enabled tool that evaluates the OCL expressions upon the previously mentioned instantiation.

Our proposal for metrics definition and collection combines formality, understandability and collection efficiency due to the usage of OCL. Furthermore, it ensures their portability among OCL-enabled CASE tools. As OCL is part of the new UML standard, an increasing number of UML CASE tools are supporting it. To collect the metrics, we execute their corresponding OCL definition upon the referred metamodel, instantiated with meta-objects representing the component assemblies to be analyzed.

Further details about this technique and case studies that illustrate its applicability have been published in the recent years. The idea of using OCL for defining software metrics to evaluate object-oriented design was proposed in [34]. FLAME, a library of OCL functions to aid in the definition and extraction of software metrics, based on the UML metamodel was presented in [43]. [44] moves to the evaluation of software components and discusses the formalization of Washizaki's metrics set [26], using the UML 2 metamodel. This formalization uses the new abstractions for software components provided by UML 2. [27] presents a case study to assess Washizaki's metrics set in a quantitative way.

Last, but not the least, the adherence to a common set of research reporting guidelines for presenting the results of experimental work in software engineering would certainly increase the comparability of different research efforts. Using research reporting guidelines such as those proposed in [45] is a promising path towards a more effective research on metrics-based approaches to software development, in general, and CBD, in particular. Note that while such guidelines are aimed at the description of experimental work (e.g. in a paper describing a controlled experiment, or set of experiments), our framework was developed to facilitate a systematic comparison of proposals found in the literature.

# 6 CONCLUSIONS AND FURTHER WORK

With the increasing demand of the software industry to include third party reusable components in the software development process, component assemblers need effective ways of selecting adequate components. Comparative reviews of existing approaches to software component evaluation are required to aid component assemblers to identify the evaluation approaches better suited to support their activity. As far as we know, this is the first attempt to provide such a review in a systematic way.

We contribute with a common framework for the characterization of component assessment proposals. The framework includes the proposal's scope, intent and used technique, a critical appreciation of the most noticeable features of the proposal, and an assessment of its maturity level, regarding the underlying quality model, its mapping to metrics, the metrics definition and the achieved level of validation.

Common problems on current approaches to CBD evaluation are identified. Overall, there is a lack of maturity in existing proposals, which is likely due to the relative novelty of black-box software components evaluation as a research topic. For instance, determining the relevant quality attributes which should be assessed is still an open issue. Ambiguity in definition of quality models and metrics, lack of adequacy of specifying formalisms and insufficient validation of proposals are among the most common shortcomings in the analysed proposals. We briefly outline our approach to mitigate these problems.

Our analysis focuses on metrics-based evaluation of structural properties of components and component assemblies. This is only part of the "evaluation toolset" that should be available to component assemblers. Other orthogonal evaluation techniques, such as the evaluation of non-functional properties and, obviously, the assessment of the composability of candidate components are essential to component selection, but are beyond the scope of the review performed in this paper.

While conducting this survey we noticed that, on other research areas, such as clinical research, where evidence-based healthcare databases are maintained (http://www.cochrane.org/), the method for gathering the proposals to be reviewed is commonly presented using clear and reproducible search and eligibility criteria. Unfortunately, this is not the current practice in the realm of Experimental Software Engineering, due to the diversity and scattering of information.

Research networks such as the ESERNET (Experimental Software Engineering Network) have attempted to coordinate efforts to mitigate the typically low number of related experimental work, as well as their diversity to an extent that makes their comparison a very hard nut to crack [46]. Empirical software engineering case studies are often conducted in an ad-hoc fashion. Striving for well-defined protocols in metrics collection experiments would significantly improve the comparability between different proposals, facilitating the production of systematic reviews upon which meta-analysis of the experimental data collected by independent research teams is more feasible.